# Mechanisms limiting the coherence time of spontaneous magnetic oscillations driven by DC spin-polarized currents

J. C. Sankey, I. N. Krivorotov, S. I. Kiselev, P. M. Braganca, N. C. Emley, R. A. Buhrman, and D. C. Ralph

*Cornell University, Ithaca, NY, 14853 USA*

The spin-transfer torque from a DC spin-polarized current can generate highly-coherent magnetic precession in nanoscale magnetic-multilayer devices. By measuring linewidths of spectra from the resulting resistance oscillations, we argue that the coherence time can be limited at low temperature by thermal deflections about the equilibrium magnetic trajectory, and at high temperature by thermally-activated transitions between dynamical modes. Surprisingly, the coherence time can be longer than predicted by simple macrospin simulations.

# I. INTRODUCTION

Recent experiments have shown that a spin-polarized DC current can excite periodic oscillations in nm-scale magnetic multilayers even in the absence of any external oscillatory drive[1-4] in agreement with predictions.[5,6] The magnetic motions produce variations of the resistance $R(t)$ that, when measured with a spectrum analyzer, give peaks in the microwave power spectral density vs. frequency (Fig. 1(a)). Deviations from perfect periodicity can be characterized by the time scale over which the oscillations lose phase coherence, related to the reciprocal of the linewidth. This scale is important both for a fundamental understanding of the dynamics and for applications including tunable nanoscale microwave sources and resonators.[7] The coherence quality has varied in previous experiments, with room-temperature linewidths ranging from a full-width-at-half-maximum (FWHM) of 550 MHz for Co layers in "nanopillars"[1] to 2 MHz for Py ($Ni_{81}Fe_{19}$) films in point-contact devices.[2] Here we investigate the processes that limit the coherence time of spin-transfer-driven precession by measuring the temperature dependence of the linewidths. We argue that two fundamental mechanisms contribute: thermal deflection of the magnetic dynamics about the equilibrium trajectory and thermally-activated transitions between dynamical modes.

# II. SAMPLE GEOMETRY

We focus on devices having a nanopillar geometry (Fig. 1(a) inset). The samples are composed of metal multilayers fabricated into elliptical cross-sections using the procedure of Ref. 1. The devices that we examine have different sequences of layers (noted below), but all contain one thin Py "free" layer (2-7 nm thick) that can be driven into precession by spin-transfer torques and a thicker or exchange-biased "fixed" Py layer that polarizes the current and does not undergo dynamics in the current range we discuss. When biased with a DC current ($I$), motion of the free-layer magnetic moment results in a microwave signal $IR(t)$ that we measure with a spectrum analyzer. Figure 1(b) is a dynamical phase diagram for Device 1, determined from microwave measurements as in Ref. 1, with magnetic field ($H$) applied in plane along the magnetically easy axis. This device has the layer structure 80 nm Cu / 20 nm Py / 6 nm Cu / 2 nm Py / 2 nm Cu / 30 nm Pt, with an approximately elliptical cross-section of 120 nm × 60 nm, and a resistance of 6 Ω (low enough that Ohmic heating[8] is negligible above 20 K). We will consider the dynamical states near bias points corresponding to the dot in Fig. 1(b) where, as a function of increasing $I$, the sample evolves from a configuration in which the moments of the two magnetic layers are parallel (P), to a dynamical mode with small-angle precession (SD), to a mode with larger-angle precession (LD).

# III. DATA AND ANALYSIS

We find that linewidths can vary significantly between samples of similar geometry, to a greater extent than the critical currents or the other aspects of spin-transfer-driven dynamics that have been analyzed previously. The differences between samples might be associated with film roughness, partial oxidation at sample edges, or other effects. We will focus on the comparatively narrow lines. Figure 2(a) shows the measured temperature dependence of the FWHM of the peak in power density observed at twice the fundamental precession frequency in Device 1.[9] Because the linewidth depends on the magnitude of the precession angle $\theta$ measured in plane (inset, Fig. 2(a)), as temperature (*T*) is changed we keep the average precession angle $\langle\theta\rangle$ approximately constant. For Device 1, we do this by monitoring the power in the second harmonic, estimating $\langle\theta\rangle$ using the procedure of Ref. 1, and adjusting *I* between 1.1 mA (25 K) and 0.9 mA (170 K) to fix $\langle\theta\rangle$ near an estimated value of 32°, where the linewidth is a minimum in this device. The misalignment angle between the precession axis and the fixed layer magnetization (estimated from the first and second harmonic[1]) is $\theta_{mis} \sim 2°$. We find that the linewidth is strongly dependent on *T*, increasing by a factor of 5 between 25 K and 170 K. We have observed qualitatively similar behavior in 6 samples, throughout the region of the phase diagram where precessional excitations exist. Figure 2(b) shows results near the fundamental precession frequency for smaller-angle precession in Device 2, composed of 80 nm Cu / 20 nm Py / 10 nm Cu / 7 nm Py / 20 nm Cu / 30 nm Pt, with cross-section 130 nm × 40 nm, and resistance 20 $\Omega$. The thicker free layer (compared to Device 1) reduces some effects of thermal fluctuations and permits studies of the small-angle dynamics up to room temperature. Measurements at the fundamental precession frequency are possible even for small $\langle\theta\rangle$ in Device 2, because of a larger value of the offset angle $\theta_{mis}$ than in Device 1. (In this case we control $\langle\theta\rangle$ by monitoring the power at the fundamental, and estimate $\langle\theta\rangle < 12°$.) The strong *T* dependence that we observe in all samples indicates that thermal effects determine the coherence time of spin-transfer-driven precession above 25 K.

To analyze these results, we first consider the simplest model, in which the moment of the free layer is assumed to respond as a single macrospin. Theoretical studies of this model have been performed previously[10-15] and good qualitative agreement has been found with both frequency-domain and time-domain measurements, with some exceptions at large currents.[1,3,4,16] We integrate the Landau-Lifshitz-Gilbert (LLG) equation of motion with the Slonczewski form of the spin-transfer torque.[17] Thermal effects are modeled by a randomly fluctuating field $\mu_0 H_{th}$, with each spatial component drawn from a Gaussian distribution of zero mean and standard deviation $\sqrt{2\alpha k_B T / \gamma M_s V \Delta t}$, where $\alpha$ is the Gilbert damping parameter, $k_B$ is Boltzmann's constant, $\gamma$ is the gyromagnetic ratio, $M_s$ and *V* are the magnetization and volume of the free layer, and $\Delta t$ is the integration time step.[18,19] Thermal fluctuations displace the moment both (a) along and (b) transverse to the equilibrium trajectory. Fluctuations along the trajectory speed and slow the moment's progress, directly inducing a spread in precession frequency *f*. From the time needed for this random-walk process to produce dephasing, we estimate the contribution to the FWHM from mechanism (a) to be

$$\Delta f_a \approx \frac{4\pi\gamma\alpha k_B T}{M_s V D^2} n^2, \qquad (1)$$

where *D* is the length of the precession trajectory on the unit sphere, and *n* = 1 or 2 for the first or second harmonic peak. If we substitute parameters appropriate for Device 1: $\alpha = 0.025$,[4] *T* = 150 K, $\mu_0 M_S$ = 0.81 T,[8] *n* = 2, dimensions 2 × 120 × 60 nm$^3$, and $\theta$ =

32°, we predict a contribution from this mechanism of $\Delta f_a \approx 12$ MHz. This is much less than the measured linewidths at $T = 150$ K, and the linear $T$ dependence also differs from the experiment, so we conclude that the contribution from this mechanism is likely negligible for Devices 1 and 2. The second mechanism, (b) thermal fluctuations of the free-layer moment transverse to the trajectory, will produce fluctuations in $\theta$ about $\langle\theta\rangle$ (upper inset, Fig. 3). If $f$ depends on $\theta$, this will cause an additional spread $\Delta f_b$. Different regimes are possible for the resulting linewidths, depending on the magnitude of $df/d\theta$, the width of the distribution in $\theta$, and the correlation time for fluctuations. However, our simulations suggest that our data correspond to a simple regime (long correlation times) in which the linewidth is simply proportional to the FWHM $\Delta\theta$ of the distribution of precession angles weighted by the magnitude of the resistance oscillations associated with each value of $\theta$:[20]

$$\Delta f_b = n \left.\frac{df}{d\theta}\right|_{\langle\theta\rangle} \Delta\theta . \qquad (2)$$

The triangles in Fig. 3 display values of the right-hand side of Eq. (2) with $\Delta\theta$ and $df/d\theta|_{\langle\theta\rangle} \approx 35$ MHz/degree both determined from the simulation. The parameters used are those corresponding to Device 1 (listed above), together with an in-plane uniaxial anisotropy $\mu_0 H_k = 20$ mT, an out-of-plane anisotropy $\mu_0 M_{eff} = 0.8$ T,[8] $I = 1.2$ mA, and $\mu_0 H = 50$ mT applied along the easy axis, with the fixed layer moment in the same direction. We assume that the angular dependence of the Slonczewski torque is simply proportional to $\sin(\theta)$ with an efficiency parameter of 0.2.[14] The squares in Fig. 3 show the FWHM calculated directly from the Fourier transform of $R(t)$ obtained in the same simulation. The agreement in Fig. 3 demonstrates that Eq. (2) gives a good description of the linewidths expected from dynamics within this approximation. The $T$ dependence of the calculated linewidths in Fig. 3 is to good accuracy $T^{1/2}$ at low $T$ (inset, Fig. 3). We expect that this form is very general (perhaps applicable even beyond the macrospin case) because this form follows from Eq. (2) if one assumes that Boltzmann statistics can be applied to this non-equilibrium problem. If fluctuations of $\theta$ about $\langle\theta\rangle$ are subject to an effective linear restoring term, then both simulations and simple analytical calculations show that $\Delta\theta \approx A\,T^{1/2}$, where $A$ is a constant.

Consider now the data for Device 1 shown in Fig. 2(a). In the range 25 - 110 K, Eq. (2) with $\Delta\theta \approx AT^{1/2}$ gives a reasonable fit, with one adjustable parameter $A\,df/d\theta|_{\langle\theta\rangle} = 2.3$ MHz K$^{-1/2}$. However, the measured widths are approximately a factor of eight narrower than those predicted by the macrospin simulation with parameters chosen to model this sample (Fig. 3). The measured value $df/d\theta|_{\langle\theta\rangle} \sim 30$ MHz/degree is similar to the simulation, so the effective linear restoring term required to model our device ($\propto 1/A^2$) would have to be larger by a factor of ~ 50. We have not been able to account for so large a difference by varying device parameters over a reasonable range or by employing different predictions for the angular dependence of the spin torque.[15] We are therefore led to the surprising suggestion that spin-transfer-driven dynamical modes can generate narrower linewidths at low $T$ than are expected within the macrospin approximation. Initial micromagnetic simulations have been performed in an attempt too account for the possibility of spatially non-uniform magnetizations in spin-transfer devices.[21-24] However, for the cases analyzed, non-uniformities have thus far led to much broader, not narrower, linewidths. It is possible that the simulations might be improved by including recently-proposed mechanisms whereby different regions of a nanomagnet interact through feedback mediated by the current.[25-28]

Above $T \approx 120$ K, the measured linewidths (Fig. 2) increase with $T$ much more rapidly than the approximate $T^{1/2}$ dependence predicted by the macrospin LLG model. A plausible mechanism for the strong $T$ dependence is switching between different dynamical magnetic modes, leading to linewidths inversely proportional to the lifetime of the precessional state. Switching between steady-state precessional modes and static states has previously been identified at frequencies from < 100 kHz[8,29,30] to 2 GHz.[31] The consequences on linewidths have been considered within LLG simulations.[14] To estimate the effects of such switching, we assume that the average lifetime of a precessional state is thermally activated, $\tau \approx (1/f) \exp(E_b / k_B T)$, where $E_b$ is an effective activation barrier. The Fourier transform then yields a linewidth

$$\Delta f_{sw} = \frac{1}{\pi \tau} = \frac{f}{\pi} \exp(-E_b / k_B T). \tag{3}$$

We find that the combination of Eqs. (2) and (3) gives a good description of the strong $T$ dependence of the linewidths in Fig. 2(a) and 2(b) using the fitting parameter $E_b/k_B = 400$ K for Device 1 and 880 K for Device 2, similar to values determined from GHz-rate telegraph-noise signals by Pufall et al.[31] Note that Eq. (3) alone would predict low-$T$ linewidths much smaller than we measure. The effective barriers from the fits are small compared to the static anisotropy barrier $\mu_0 M_s H_k V / k_B$ ~ 10,000 K in Device 1 and 100,000 K in Device 2. It is not surprising that the effective barriers for switching between dynamical states are distinct from the static anisotropies.

Direct evidence for the importance of the switching mechanism can be seen in some samples (e.g., Device 3, composition 80 nm Cu / 8 nm IrMn / 4 nm Py / 8 nm Cu / 4 nm Py / 20 nm Cu / 30 nm Pt, with a cross section 130 × 60 nm$^2$) for which, at particular values of $I$, $H$, and $T$, multiple peaks can appear simultaneously in the power spectrum at frequencies that are not related harmonically (Fig. 4). In these regimes, the widths of both peaks are broader than when only a single mode is visible in the spectrum. We suggest the cause is rapid switching between two different dynamical states.

Macrospin simulations at experimental temperatures do not exhibit high-frequency switching between metastable states except in narrow regions of the dynamical phase diagram where nearly-degenerate modes exist.[12,14] In contrast, we measure strong thermally-activated temperature dependence whenever precessional dynamics are present, for $T > 120$ K. In this regime, transitions involving non-uniform modes[22,25,26] therefore appear only to increase the linewidths. Understanding these transitions will provide an important test for future micromagnetic simulations.

The narrowest linewidth that we have achieved for free-layer oscillations in a nanopillar device is 5.2 MHz (Fig. 1(a), for a sample composition the same as Device 3), corresponding to a coherence time $1/\Delta f$ ~ 190 ns. This is more than a factor of 100 improvement relative to the first measurements in nanopillars,[1] and is comparable to the limit expected from Eq. (1).[32] Such narrow linewidths are observed in devices containing an antiferromagnetic layer to exchange-bias the fixed magnetic layer 45° relative to the easy axis and with *H* applied along the exchange-bias direction. We speculate that the reduced symmetry of these conditions may improve the coherence time by reducing both *df/dθ* and the likelihood of thermally-activated switching between low-energy modes.

## IV. CONCLUSIONS

In summary, our data indicate that the coherence time of spontaneous spin-transfer-driven magnetic dynamics is limited by thermal effects: thermal fluctuations of the precession angle at low *T*, and thermally-activated mode switching at high *T* or near bias points where two or more different modes are accessible. The coherence time can be increased dramatically by cooling samples below room temperature.

We thank J. Xiao and M. Stiles for helpful discussions. We acknowledge support from DARPA through Motorola, from the Army Research Office, and from the NSF/NSEC program through the Cornell Center for Nanoscale Systems. We also acknowledge NSF support through use of the Cornell Nanofabrication Facility/NNIN and the Cornell Center for Materials Research facilities.

**FIGURES**

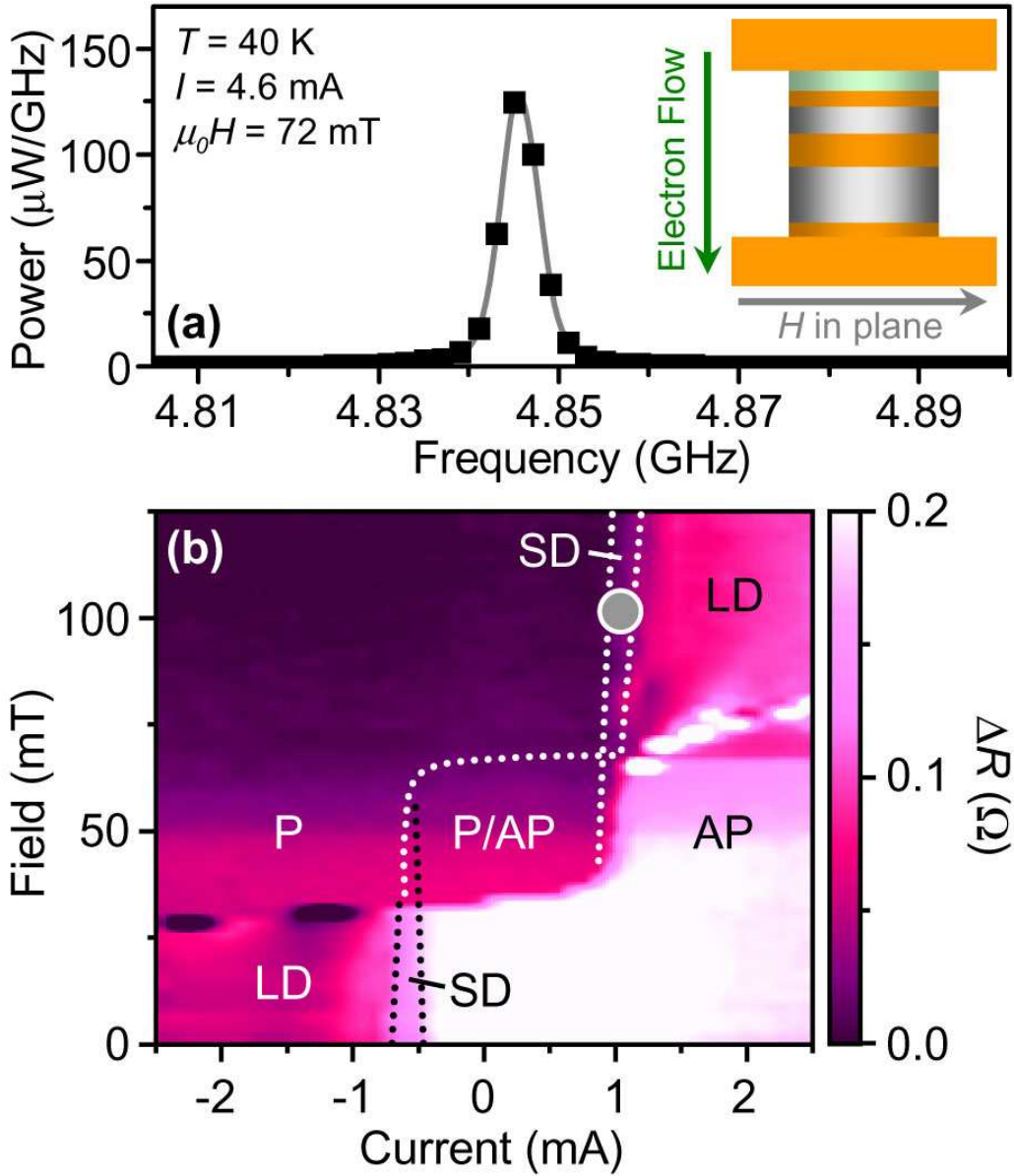

FIG. 1 (Color online) (a) The narrowest spectral peak for free-layer oscillations that we have observed in a nanopillar (FWHM = 5.2 MHz). The device has the same composition as Device 3, described in the text. (inset) Schematic of a nanopillar device. (b) Differential resistance of Device 1 as a function of $I$ and $H$ at $T = 4.2$ K, obtained by increasing $I$ at fixed $H$. AP denotes static antiparallel alignment of the two magnetic moments, P parallel alignment, P/AP a bistable region, SD small-angle dynamics, and LD large-angle dynamics.

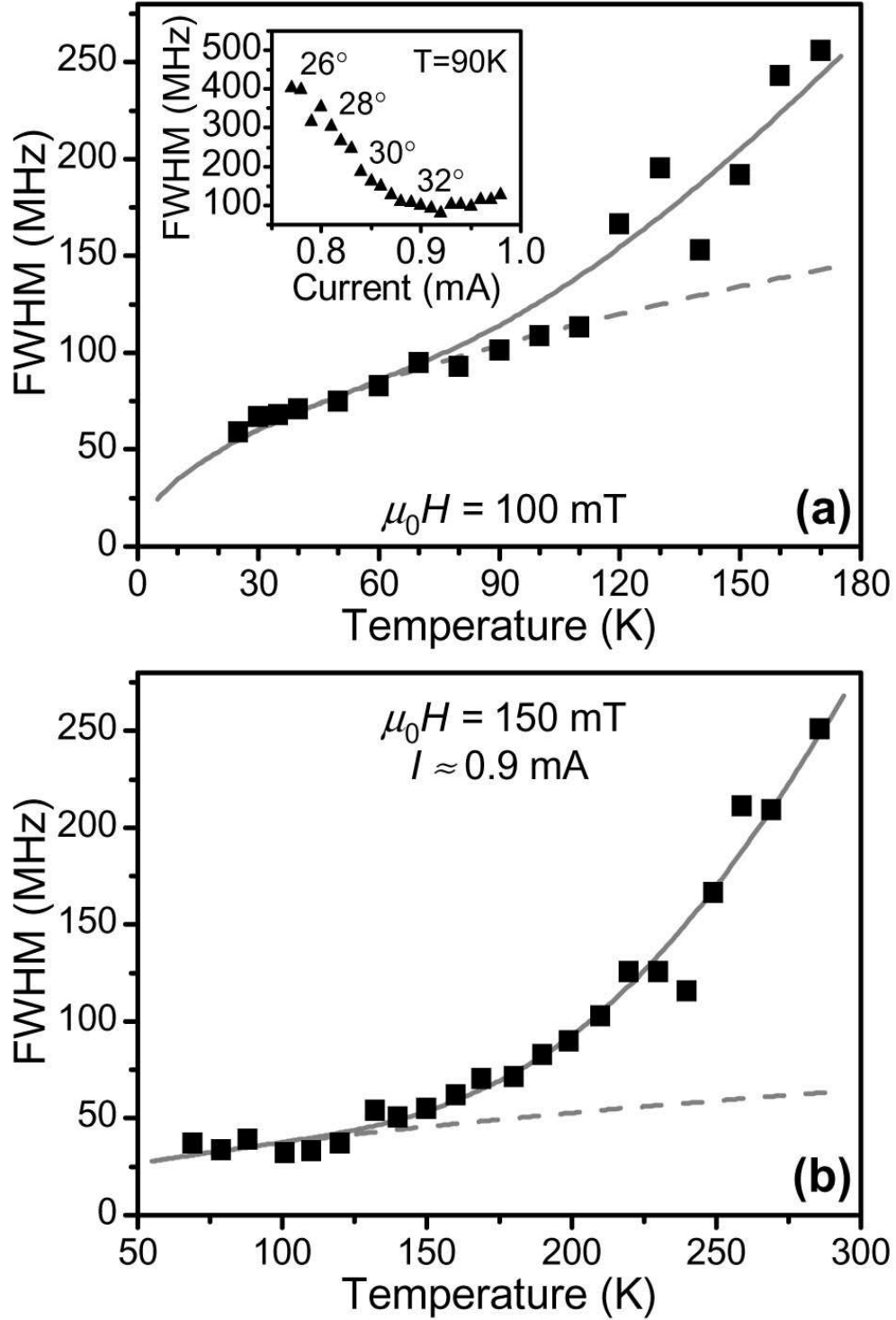

FIG. 2 Measured linewidths vs. *T* for (a) Device 1 and (b) Device 2. The dashed line is a fit of the low-*T* data to Eq. (2) and the solid line is a combined linewidth from Eqs. (2) and (3), obtained by convolution. (inset) Dependence of linewidth on *I* for Device 1, with estimates of precession angles.

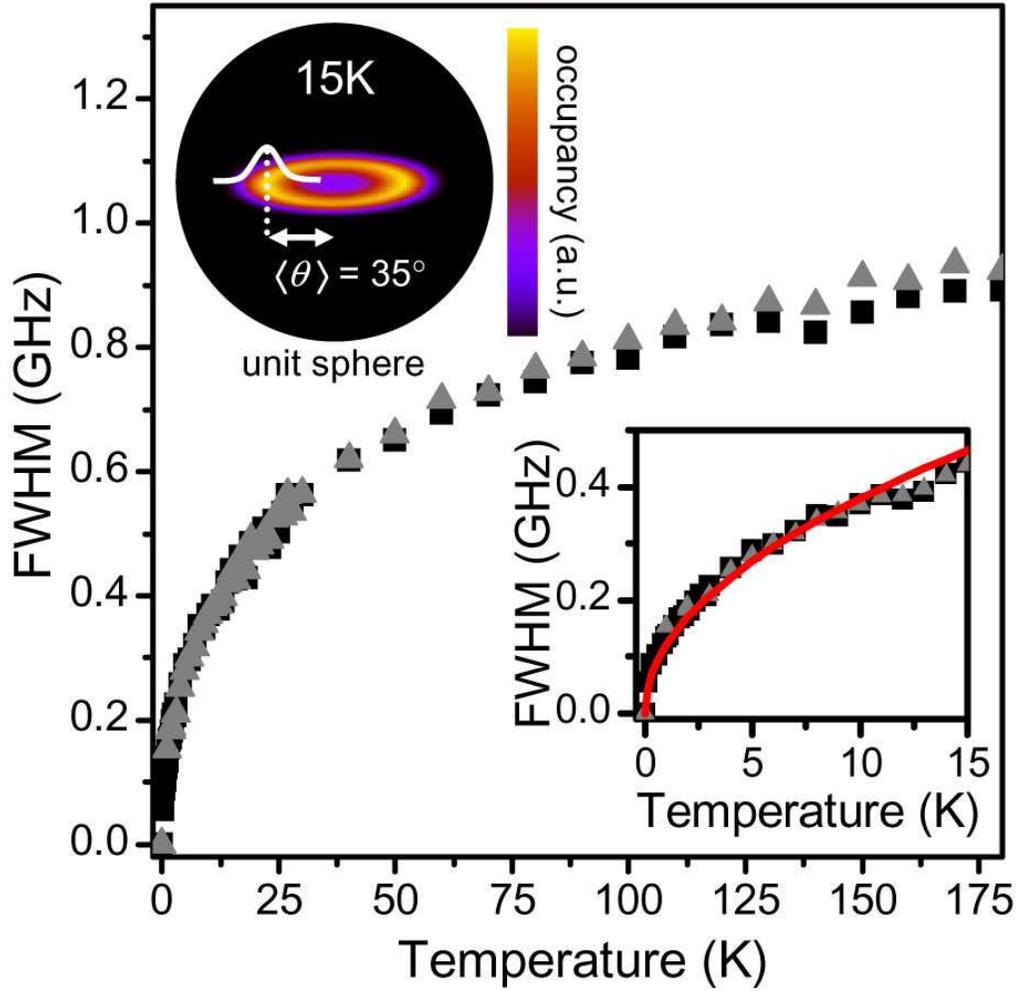

FIG. 3 (Color online) (Main plot and lower inset) Squares: Linewidth calculated directly from the Fourier transform of $R(t)$ within a macrospin LLG simulation of the dynamics of Device 1. Triangles: Linewidth calculated from the same simulation using the right-hand side of Eq. (2). Line in inset: Fit to a $T^{1/2}$ dependence. (Top inset) Simulated probability distribution of the precession angle at 15 K.

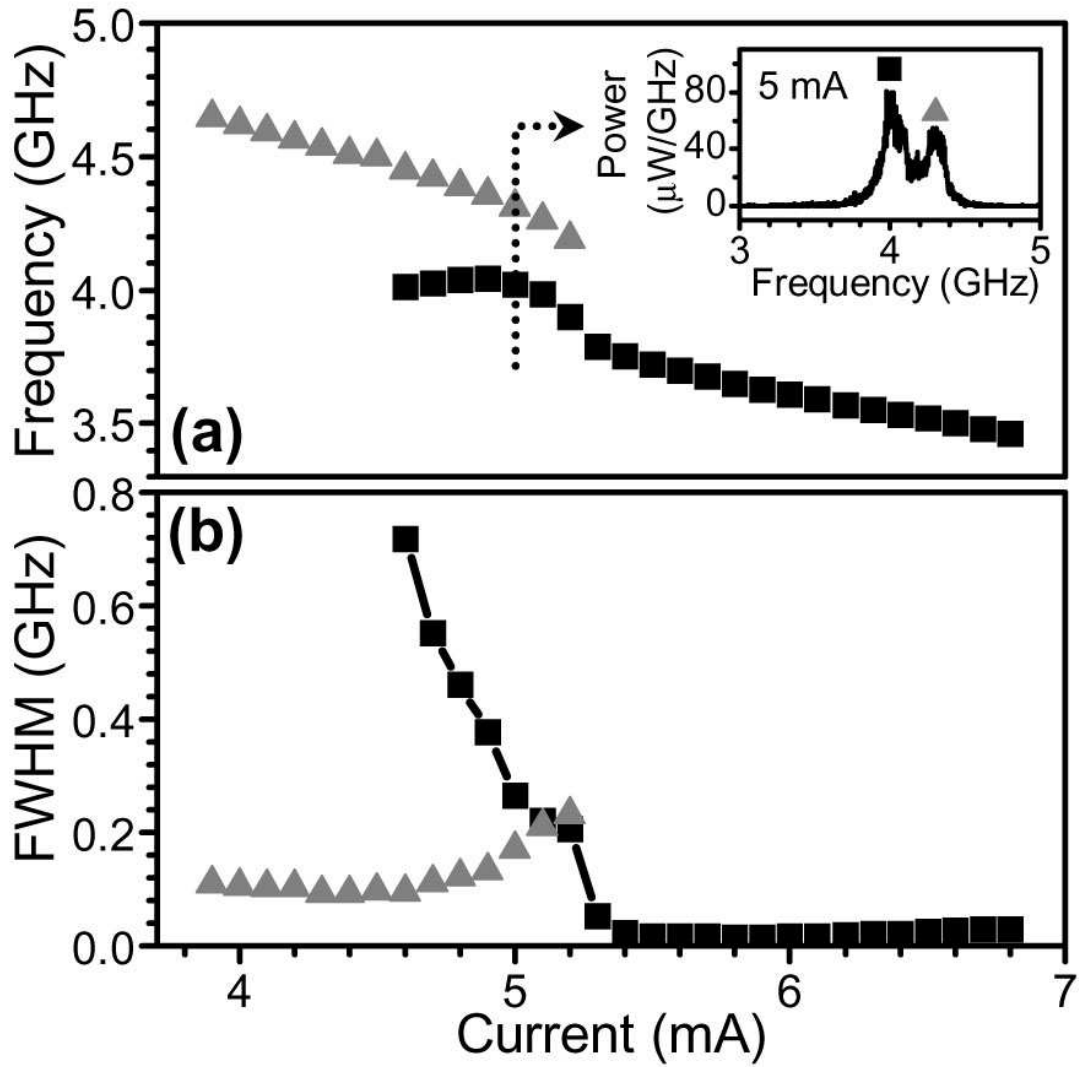

FIG. 4 Measured (a) frequencies and (b) linewidths of large-angle dynamical modes in Device 3 for $T = 40$ K, $\mu_0 H = 63.5$ mT applied in the exchange bias direction, 45° from the free-layer easy axis. When two modes are observed in the spectrum simultaneously, both linewidths increase.

32 Ohmic heating due to increased critical currents and resistance precludes $T$ dependendent studies below $T \approx 200$ K in the samples with our smallest linewidths.